# Well Cement Degradation and Wellbore Integrity in Geological CO$_2$ Storages: A Literature Review

## Nguyen V*, Olatunji O, Guo B and Ning Liu

Department of Petroleum Engineering, University of Louisiana at Lafayette, USA

**\*Corresponding author:** Nguyen V, Department of Petroleum Engineering, University of Louisiana at Lafayette, LA 70504, USA, Tel: 6787903709; Email: vu.nguyen1@louisiana.edu

## Abstract

Carbon capture and storage (CCS) has emerged as the most effective method to curb the CO$_2$ concentration in the atmosphere. It can store up to 5 billion tons of CO$_2$ per year. To guarantee a safe and economical geological storage, the well cement degradation and wellbore integrity need to be studied thoroughly. This review paper is designed to provide a fundamental background of well cement degradation and wellbore integrity in geological CO$_2$ storages to support the researchers in further investigation. The review mainly focuses on mechanical, thermal, chemical property changes and corrosion time for cement in experiments and simulation during geological CO$_2$ storage. However, the debonding interface between casing/cement or cement/formation has not been addressed profoundly. A further investigation should inspect how pressure, temperature, and chemical reaction affect the micro-annuli of casing/cement or cement/formation. Also, a mathematical model should be established to predict the corrosion rate in geological CO$_2$ storage.

## Introduction

Portland cement is composed of four major components: tricalcium silicate (Ca$_3$SiO$_5$ or C$_3$S), dicalcium silicate (Ca$_2$SiO$_4$ or C$_2$S), tricalcium aluminate (Ca$_3$Al$_2$O$_6$ or C$_3$A), tetracalcium aluminoferrite (Ca$_4$Al$_2$Fe$_2$O$_{10}$ or C$_4$AF) [1]. The composition of Portland cement [2] can be found in Table 1. Corresponding to MacLaren DC [3], the Portland cement dissolves in water through the following reactions in Equation1 and Equation 2:

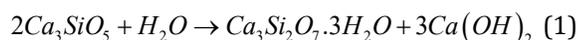

$$2Ca_3SiO_5 + H_2O \rightarrow Ca_3Si_2O_7.3H_2O + 3Ca(OH)_2 \text{ (1)}$$

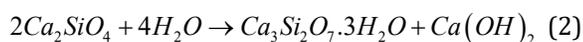

$$2Ca_2SiO_4 + 4H_2O \rightarrow Ca_3Si_2O_7.3H_2O + Ca(OH)_2 \text{ (2)}$$

An understanding of cement corrosion during the geologic storage of CO$_2$ is critical. The mechanism of cement corrosion takes place in the following sequence of reaction:

CO$_2$ will first dissolve in Calcium hydroxide (CH) to form calcite [4] ( calcium carbonate) by the reaction in Equation 3:

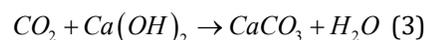

$$CO_2 + Ca(OH)_2 \rightarrow CaCO_3 + H_2O \text{ (3)}$$

This reaction benefits the mechanical property of the wellbore. Specifically, calcite would enhance the mechanical strength of the cement layer and reduce porosity, which results in the permeability decreasing. The decline of permeability characteristics will retard the possibility of CO$_2$ leakage.

Then the Calcium Silicate Hydrate (Ca$_3$SiO$_7$.3H$_2$O or C-H-S) in cement reacts with CO$_2$ by Equation 4

$$Ca_3Si_2O_7.3H_2O + 3CO_2 \rightarrow 3CaCO_3 + 2SiO_2 + H_2O \quad (4)$$

Cement quality is reduced when it is converted to bicarbonate in the presence of excess $CO_2$ [5] in Equation 5.

$$CO_2 + CaCO_3 + H_2O \rightarrow Ca(HCO_3)_2 \quad (5)$$

Dissolution of the bicarbonate in the presence of water leaves the material more porous. This reaction is a great issue to stop $CO_2$ from leaking. The challenge is how to control the bicarbonate reaction and predict the penetration depth of carbonation to protect the cement layer sheath from damage. Several groups have been conducting experiments to enhance the trait of cement to inhibit the corrosion process [6-10]. Others used the mathematical model to simulate how quickly $CO_2$ would penetrate by diffusion mechanism through the cement layer [11,12]. Some groups carried out an experiment by measuring the penetration depth versus time with support from techniques such as Secondary Energy Microscopy, X-Ray Diffraction, Scanning Electron Microscope (SEM)-Back Scattered Electrons (BSE) micrograph, and Energy Dispersive Spectroscopy (EDS) map, etc [13,14]. The penetration depth was found as a function of the square root of time. However, these studies are time consuming, expensive, and prone to error. Therefore, an accurate mathematical model is required to govern the diffusion process to reduce any drawback by experiment-induced and increase the answer's reliability. The diffusion equation (Equation 6) by Fick's second law will suit this requisition:

$$\partial C / \partial t = D(\partial^2 C) / (\partial^2 x^2) \quad (6)$$

Where C is the carbonate concentration; t is the cement exposure time, D is the diffusion coefficient; x is the penetration depth.

| ASTM type | Description | $C_3S$ | $C_2S$ | $C_3A$ | $C_3AF$ | $CS_2$ |
|-----------|-------------|--------|--------|--------|---------|--------|
| I | General purpose | 55 | 17 | 10 | 7 | 6 |
| II | Moderate sulfate resistant | 55 | 20 | 6 | 10 | 5 |
| III | High early strength | 55 | 17 | 9 | 8 | 7 |
| IV | Low heat of hydration | 35 | 40 | 4 | 12 | 4 |
| V | Sulfate resistant | 55 | 20 | 4 | 12 | 4 |

*(C=Cao, S=SiO$_2$, A=Al$_2$O$_3$, and F= Fe$_2$O$_2$)
**Table 1:** Composition of Portland cement (wt.%).

## Cement and Wellbore Property

One of the crucial tasks of a wellbore is well cementing. The purposes of well cementing are to reinforce the casing strength, prevent undesirable formations away from the borehole, inhibit casing corrosion rate, constraint irregular pore pressure, etc. The most substantial function of cementing is to obtain zonal isolation. The other duty of cementing that is also important is to acquire a good bond between cement and pipe or cement and formation. However, the cement characteristics are changed over time. During CO2 storage, the three most common changes in the cement and in the rock that need to be concerned are mechanical, thermal, and chemical properties.

### Mechanical Property

The pressure build-up due to $CO_2$ injection may result in the poro-mechanical effect. The change in pressure leads to a change in the stress field, which causes a mechanical impact on the fault and top seals. The mechanical defect on seals and faults is a reason not only for the migration of $CO_2$ out of the well but also for leading ground movement.

The most prevalent criteria to evaluate the failure of rock and cement in geomechanics is Mohr-Coulomb. This hypothesis assessed the failure of material based on the combination of normal and shear stresses. The failure line is computed by Equation 7:

$$\tau_f = \tau_o + \sigma tan(\phi) \quad (7)$$

Where: $\tau_f$, $\tau_o$, $\sigma$ and $\phi$ are shear strength, cohesion, effective normal stress, and internal friction angle, respectively.

The normal and shear effective stresses are given in Equation 8 and Equation 9 and demonstrated in Figure 1. The shear failure occurs as $(\sigma,\tau)$ hits or crosses the failure line $\tau_f$. The tensile failure takes place as $(\sigma, \tau)$ reaches or crosses the tensile axis, $\tau$.

$$\tau = \frac{1}{2}(\sigma_1 - \sigma_3)\sin 2\beta \quad (8)$$

$$\sigma = \frac{1}{2}(\sigma_1 + \sigma_3) + \frac{1}{2}(\sigma_1 - \sigma_3)cos 2\beta \quad (9)$$

where $\sigma_1, \sigma_3$ are the maximum and minimum principal stresses, respectively.

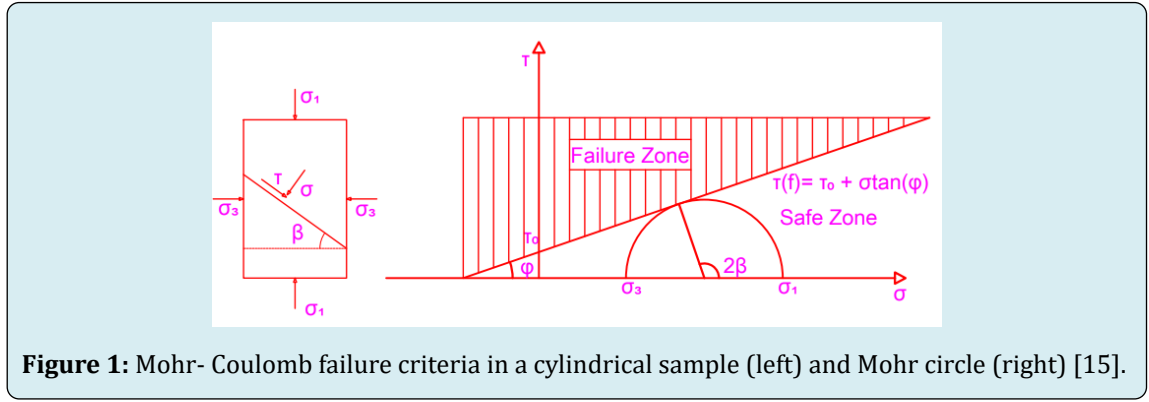

**Figure 1:** Mohr- Coulomb failure criteria in a cylindrical sample (left) and Mohr circle (right) [15].

Numerous failure mechanisms can take place in a cement sheath, such as inner debonding, outer debonding, radial cracks, shear cracks, etc. Figure 2 would interpret these mechanisms.

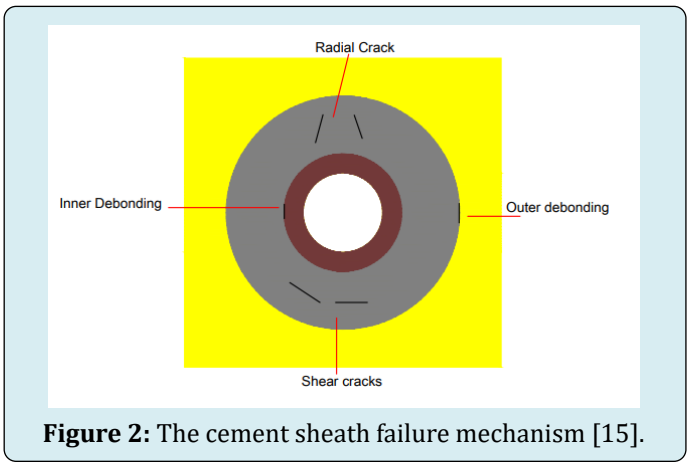

**Figure 2:** The cement sheath failure mechanism [15].

Orlic B, et al. [16] examined the poro-mechanical effects for geological $CO_2$ storage in a depleted gas field [17,18] and a deep saline aquifer in the Netherlands.

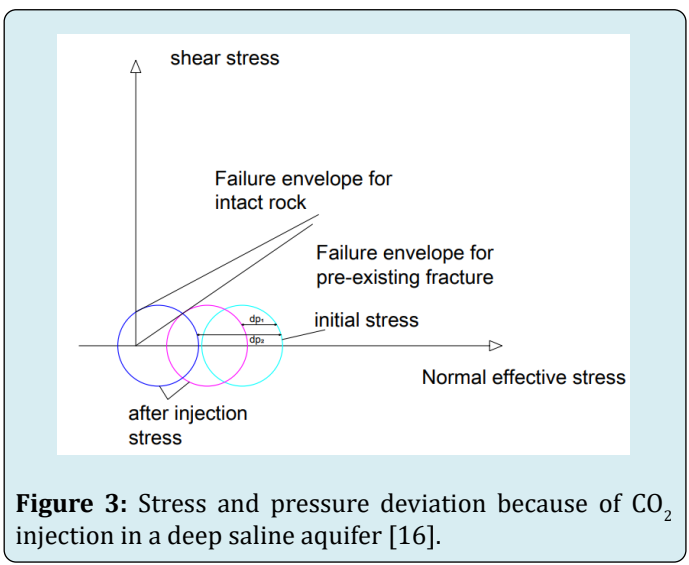

**Figure 3:** Stress and pressure deviation because of $CO_2$ injection in a deep saline aquifer [16].

Figure 3 presented the shear stress and the normal stress before and after $CO_2$-injection. The simulation was performed by increasing pressure dp1 and dp2 with dp2>dp1. As seen, the Mohr-Coulomb circle shifted more to the left with increasing pressure. It indicates that the normal and shear stress tended to reach closer to the failure line and shear stress axis. Therefore, the failure may occur easily.

Li B, et al. [19] provided a model to assess the cement sheath integrity in the carbon sequestration fields. Then the model was verified with Atkinson's model and Carter's finite element model. A case study showed that maximum injection pressure would be under-evaluated if cement sheath-induced stress is not examined. One of the conclusions is that the critical failure point lied at the cement-sandstone/shale formation interface.

Omosebi O, et al. [20] performed an experiment to investigate the cement integrity during $CO_2$ injection. An important conclusion was that class H cement turned into more elastic than class G cement with $CO_2$. The differences in composition and mineralogy can explain it.

**Thermal Property**

The changes in temperature amid $CO_2$ injection may induce stress variation, which can devastate the well. The cold injected $CO_2$ in contact with warm wellbore, reservoir and the thermal properties difference between wellbore casing, lithology, and the cement will stress the vicinity of a wellbore. As a result, the chances of creating more leakage paths are very high.

Thermal stress, which results in deterioration in cement sheath, relies on the thermo-mechanical properties of the material, the azimuthal extent, and radial position of fracture in addition to injection temperature and effective in-situ horizontal stress. The thermal-mechanical properties consist of elastic modulus, thermal conductivity, and the thermal expansion coefficient of cement. Thermal conductivity is used to measure how much heat is transmitted through the wellbore materials. The thermal expansion measure how much cement would experience shrinkage or swell due to the change in temperature.

Experimental and numerical models have been used to study the thermal effect on wellbore integrity. The experiments were performed by Shadravan A, et al. [21] by applying different pressure on the casing at high temperatures. Teodoriu C, et al. [22] designed a ring similar to a cement sample which was exposed to various internal pressure at high temperatures. Boukhelifa L, et al. [23] performed studies about sealants under different wellbore conditions, including the pressure, temperature, and geometry changes. The cracks tending to be perpendicular were observed. It is in accordance with the theory. The casing radial displacement is more prominent than its axial displacement, so the possibility for cement sheath cracking to be orthogonal with its radius is high. One of the most significant constraints is the mechanical loading corresponding to the temperature changes, which has not been fully investigated. Furthermore, the thermal property of the material was not taken into serious account. Todorovic J, et al. [24] revealed that water saturation is a critical parameter to damage wellbore integrity in harsh cooling conditions. This concludes that during the $CO_2$ injection, the possibility for cement and formation failure would surge. Aursand P, et al. [25] used a mathematical model to couple two-phase flow of $CO_2$ and radial heat transfer between the $CO_2$ flow and the well geometry. The study shows that the most considerable downhole temperature variations take place in the bottom part of the well. It also states that the parameters such as injection temperature, injection flow rate, injection duration, and downtime would affect the thermal stress leading to the damage of wellbore integrity. Lund H, et al. [26] introduced a heat-conduction model to compute the radial heat transfer from the well to the casing, annular seal, and rock formation. The model reveals that displacing cement with an annular sealant material with higher thermal conductivity would reduce the temperature variation between the casing/seal interface and the seal/rock interface. Ruan B, et al. [27] set up a two-dimensional radial wellbore flow model and solve the mass equation, momentum equations, and energy equation to investigate the thermal behavior of $CO_2$. The study showed the temperature profile along the radial and axial direction, which is crucial information to predict the thermal stresses along the casing pipe and the outer cement. Lavrov A, et al. [28] initiated the study, which investigated that the most sensitive part of tensile cracking during $CO_2$ injection is cement adjacent to the casing pipe. The study suggested that reducing stiffness and increasing the thermal conductivity of damaged materials would inhibit the number of tensile cracks.

The thermal stress is proportional to Young's modulus E(MPa), and the linear thermal expansion coefficient α; inversely proportional to Poisson's ratio ν. The relationship between them is shown in Equation 10:

$$\Delta \sigma_T = \frac{\alpha E \Delta T}{1-\nu} \quad (10)$$

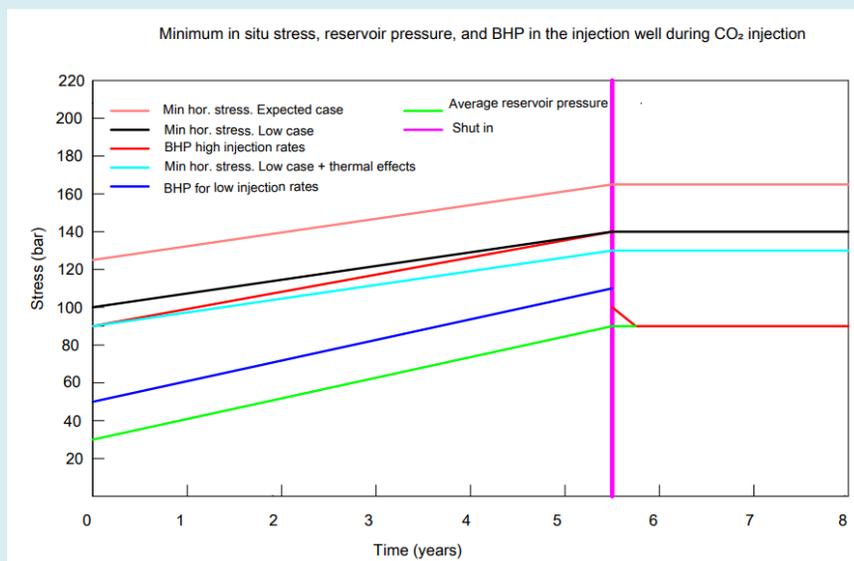

**Figure 4:** Growth of the minimum in situ stress and reservoir pressure compared to the bottom hole pressure (BHP) during $CO_2$ injection [16].

To illustrate the thermal effects on the stability of wellbore, the simulation was performed at sandstone depth of 1300m and the temperature difference of 20℃ between $CO_2$ and reservoir rock. As shown in Figure 4 for the high injection rate, the minimum horizontal stress with low case and thermal effects is lower than the bottom hole pressure. It signifies that the fracture would take place.

| Coefficient of thermal expansion | $10^{-5}$ (1/℃) |
|---|---|
| Poisson's ratio | 0.2-0.21 |
| Modulus of elasticity | 14-41 Gpa |

**Table 2:** Typical properties of Portland cement concrete [29].

The typical properties of Portland cement concrete were presented in Table 2. The values of thermal expansion coefficient and Poisson's ratio are small while big for modulus of elasticity. Therefore, based on Equation 10, reducing the cement elastic modulus is an effective way to decrease thermal stress. The relationship between thermal stress and elastic modulus was also studied thoroughly in Thiercelin, et al. [30]. This study concluded that increasing the temperature change would advance the thermal stress and a low cement elastic modulus would adapt better than a high one.

### Chemical Effect

The injection of $CO_2$ also affects rock stability in terms of chemical effects. Previous studies were carried out to demonstrate the chemical effects. One of them can be found in Orlic B, et al. [16]. The Permian Zechstein formation was selected for investigating the effect of $CO_2$ storage. The design composed of circular elements of rosettes (60% volume) in the size of millimeters implanted in a matrix of 3 types of anhydrite circular element with the size of 50-83 micrometer. The conditions for simulation are vertical stress= 50 MPa, horizontal stress=40 MPa, T=80℃ for 50,000 years. The reaction of anhydrite with $CO_2$ and water is in Equation 11:

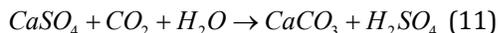

$$CaSO_4 + CO_2 + H_2O \rightarrow CaCO_3 + H_2SO_4 \quad (11)$$

According to the authors, the composition of anhydrite caprock was selected from the Permian Zechstein, which was already studied by Hangx SJT, et al. [31], in order to compare the results. The condition for vertical stress, horizontal stress, and temperature for this model has emulated the condition for caprock buried at 2.5 km depth.

The anhydrite failure strength was significantly reduced by 25% over the course of 50,000 years in a $CO_2$-rich environment. For 1000 years, the anhydrite failure strength reduction is inconsiderable. These results match with Hangx SJT, et al. [31].

For injecting $CO_2$ into a deep aquifer and depleted oil reservoir, $CO_2$ dissolves into brine to create carbonic acid, $H_2CO_3$, then reacts with rock formation (carbonate). The chemical reactions are presented in Equation 12 and Equation 13. The dissolution of rock by carbonic acid causes the rock properties changes such as geomechanical and petrophysical, which were investigated in Kim K, et al. [32]; Charalampidou EM, et al. [33]; Luquot L, et al. [34]; Rohmer J, et al. [35]; Bemer E, et al. [36]; Vanorio TV, et al. [37], Iyer J, et al. [38]; Dávila G, et al. [39]. These studies revealed that injecting $CO_2$ would increase the rock porosity and decrease the elastic moduli.

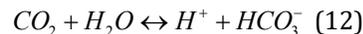

$$CO_2 + H_2O \leftrightarrow H^+ + HCO_3^- \quad (12)$$

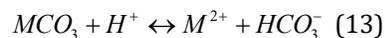

$$MCO_3 + H^+ \leftrightarrow M^{2+} + HCO_3^- \quad (13)$$

Where M is metal such as Ca, Mg.

Tang Y, et al. [40] carried the dynamic and static experiments to interpret the impact of $CO_2$-brine- rock interaction in gas reservoir with an aquifer. The results discovered that $CO_2$ brine-rock interaction takes place in both gas zone and water zone because water vaporizes into gas zone to contact with $CO_2$ to form carbonic acid. Six cores representing three reservoir types that are different in length, diameter, porosity, and permeability, were selected to perform the investigation. In general, the core porosity would be increased, and the core permeability could be decreased, as presented in Figure 5 and Figure 6. It can be demonstrated by mineral dissolution and particle migration in the pore space. Mineral dissolution induces the increasing of rock porosity, and particle migration in the pore space retards the flow resulting in the decreasing of rock permeability. However, two irregular cases were observed. The porosity of core #1 decreases, and the permeability of core #6 increases. It can be explained by the characteristics of core # 1 and # 6. The pore diameter in core #1 is small, so minerals dissolution cannot be driven out easily, causing the decreasing of porosity. In contrast, the pore size of core #6 is large, the free grains move out the pore smoothly, resulting in the decreasing of rock permeability.

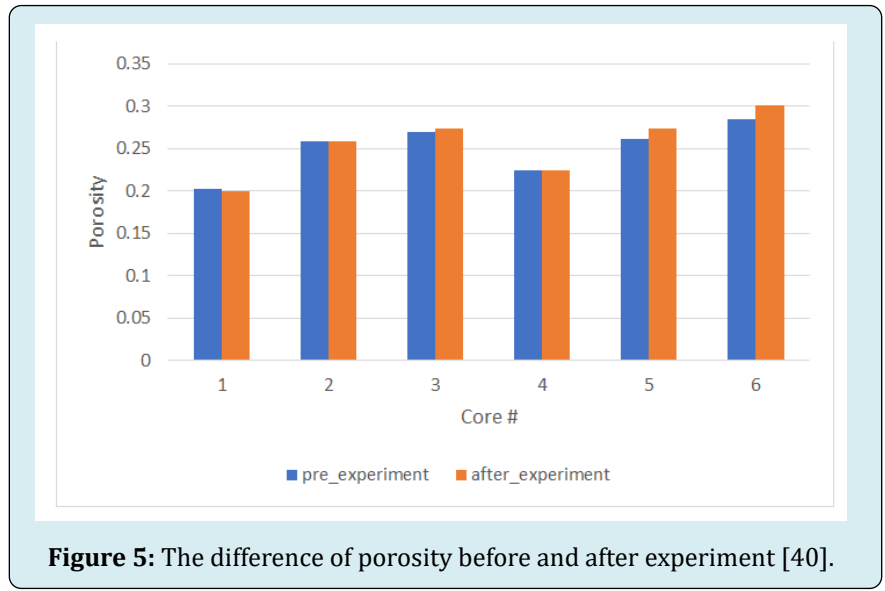

**Figure 5:** The difference of porosity before and after experiment [40].

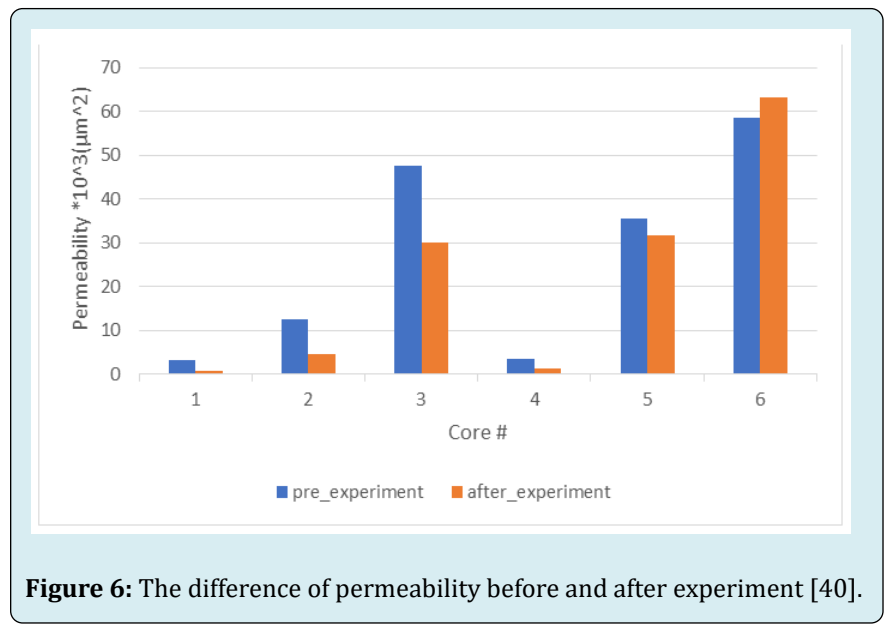

**Figure 6:** The difference of permeability before and after experiment [40].

There are still several studies Santra A, et al. [41]; Ojala IO [42]; Morris JP, et al. [43]; Karimnezhad M, et al. [44]) which were discussed in detail the $CO_2$-injection induced the changes of mechanical, chemical, and thermal property. These studies revealed that the tensile strength is a rock porosity dependence. Particularly, increasing porosity would decrease the tensile strength by an exponential function.

## Corrosion

The corrosion costs billions of dollars annually and damages our environment due to the leakage of unwanted fluid. An understanding of the process is very crucial to reduce costs and handle the process. There are two different corrosion mechanism that occur in the wellbore cement: the first due to electrochemical reaction and the second caused by the carbonation reaction [45,46].

### Electrochemical Corrosion Mechanism

This is metal corrosion. The oxidation-reduction reactions take place at anode and cathode. Particularly, the anodic reaction (Equation 14) is the oxidation of iron and the cathodic reaction (Equation 15) is hydrogen evolution.

Anode:

$$Fe \leftrightarrow Fe^{2+} + 2e^- \quad (14)$$

Cathode:

$$2H^+ + 2e^- \rightarrow H_2 \quad (15)$$

Several researchers have carried out studies to inspect the metal corrosion rate. Nevertheless, the corrosion process under high pressure of $CO_2$ (above the critical point at 7.38 MPa at 31.1ºC as seen in Figure 7) needs to be investigated further.

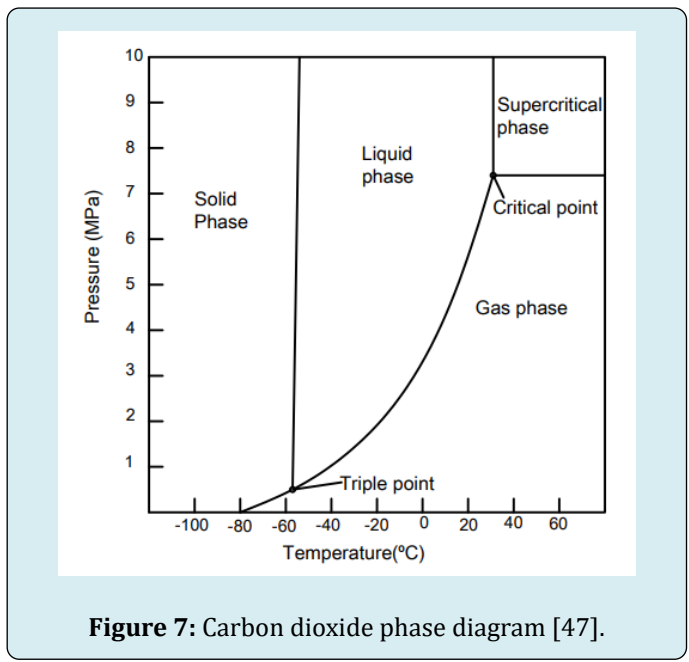

**Figure 7:** Carbon dioxide phase diagram [47].

Russick EM, et al. [48] carried the experiment on stainless steels (304L and 316), copper (CDA 101), aluminum alloys (2024, 6061, and 7075), and carbon steel (1018) in contact with pure supercritical $CO_2$, water-saturated $CO_2$, the mixture of supercritical $CO_2$ with 10 wt% of methanol, and supercritical $CO_2$ with 4 wt % of tetrahydrofurfuryl alcohol (THFA) at 3500 psi and 50 ºC. No sign of corrosion on any metal was observed when they are in contact with pure supercritical $CO_2$. For water-saturated $CO_2$, only carbon steel 1018 was sensitive, while the others were not influenced. The copper CDA 101 and aluminum 2024 got corrosive with the combination of supercritical $CO_2$ and 10 wt% of methanol. The mixture of supercritical $CO_2$ with 4 wt% of THFA almost did not cause corrosion on any metal. The THFA comprise an organic additive, Polygard, which acts like a corrosion inhibitor.

Seiersten M, et al. [49] studied the impact of the pressure of $CO_2$ up to 80 bar and temperature up to 50ºC to corrosion rate of carbon steel X65. The study presented that dry $CO_2$ and non-$CO_2$ saturated with water did not cause corrosion to carbon steel. At 50ºC in the systems consisting of only water, the corrosion rate positively correlates with $CO_2$ partial pressure. The corrosion rate reaches the maximum value of 6.9 mm/year at 40 bar. Seiersten M [50] revealed that at 4 ºC, increasing the $CO_2$ partial pressure would decrease the corrosion rate. The maximum value for corrosion rate is about 5.6 mm/year at 10 bar. The difference between the two observations above can be demonstrated by the formation of the film $FeCO_3$. At 40ºC, increasing $CO_2$ partial pressure would decrease the saturation, which leads to non-creation of film. In contrast, at 50 ºC the solution saturation has a positive correlation with $CO_2$ partial pressure.

Choi YS, et al. [51] investigated the behavior of carbon steel under the $CO_2$- saturated water phase and the water-saturated $CO_2$ phase with and without the presence of oxygen. The research exhibited that oxygen would make the corrosion rate faster. The presence of oxygen would inhibit the formation of the defensive film layer $FeCO_3$, which leads to an increase in the corrosion rate. The increasing corrosion rate with the presence of oxygen is also demonstrated by the oxidation-reduction reaction mechanism. Oxygen acts as an oxidizing agent, which would make the redox reaction between iron, oxygen, and water happen. The SEM and EDS techniques also were conducted to justify the results.

Lin G, et al. [52] examined the influence of $CO_2$ at different temperatures and pressure in autoclaves on three types of carbon steel N80, P110, and J55. At 6.89 MPa and 90ºC, the corrosion rate are 1.752 mm/y, 2.403 mm/y, and 1.854 mm/y for N80, P110, an J55, respectively. On the other hand, at higher pressure 10.34 MPa and 90 oC, those values seem likely to decrease. They are 0.922 mm/y, 1.054 mm/y, 1.105 mm/y. All values were plotted in Figure 8. The average decreasing percentage of corrosion rate for N80, P110, and J55 is 95 %, and the most decreasing corrosion rate is for P110 with 127%. P110 is also the most corrosive steel in this study. The illustration to explain the most corrosive characteristic of P110 lies in its composition as presented in Table 3. Steel P110 contains the most manganese among three types of steel N80, P110, and J55, and manganese is a very strong oxidizing agent. Therefore, the corrosion rate on P110 is almost higher comparing to N80 and J55.

| Steel Grade | C | Si | Mn | P | S | Cr | Mo | Ni | Ti | Cu |
|---|---|---|---|---|---|---|---|---|---|---|
| N80 | 0.24 | 0.22 | 1.19 | 0.013 | 0.004 | 0.036 | 0.021 | 0.028 | 0.011 | <0.019 |
| P110 | 0.26 | 0.2 | 1.4 | 0.009 | 0.003 | 0.15 | 0.01 | 0.12 | 0.03 | <0.01 |
| J55 | 0.19 | 0.31 | 1.39 | 0.014 | 0.004 | 0.19 | 0.092 | 0.017 | 0.04 | <0.01 |

**Table 3:** Steel composition (wt%) [33].

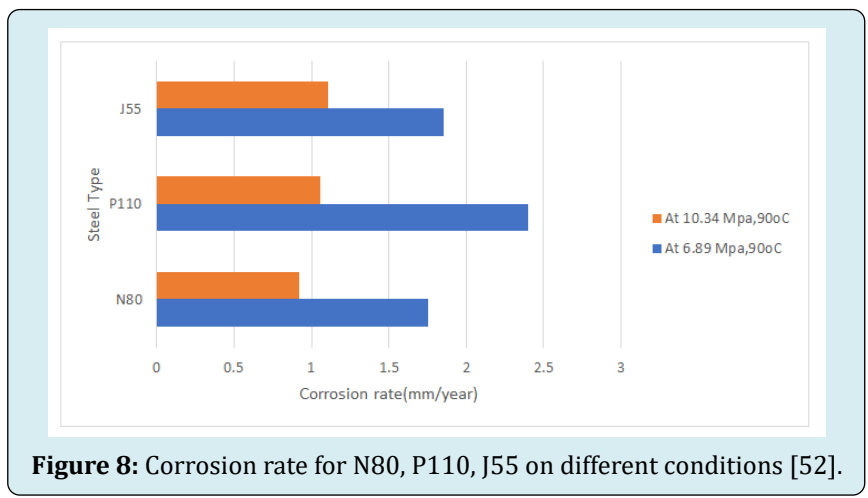

**Figure 8:** Corrosion rate for N80, P110, J55 on different conditions [52].

**Cement Corrosion Mechanism**

Understanding the cement corrosion during the $CO_2$ geological storage is necessary to handle the process correctly. This includes the $CO_2$ leakage paths and the carbonation time. Recognizing all possible leakage paths would aid in looking for a reasonable solution to cracking problems. Interpreting the carbonation time could contribute to evaluating the safety of the process for an extended period.

**$CO_2$ Leakage Path:** There are many leakages pathways for $CO_2$. It could be an interface formation-cement or cement-casing, or from cement cracking as shown in Figure 9.

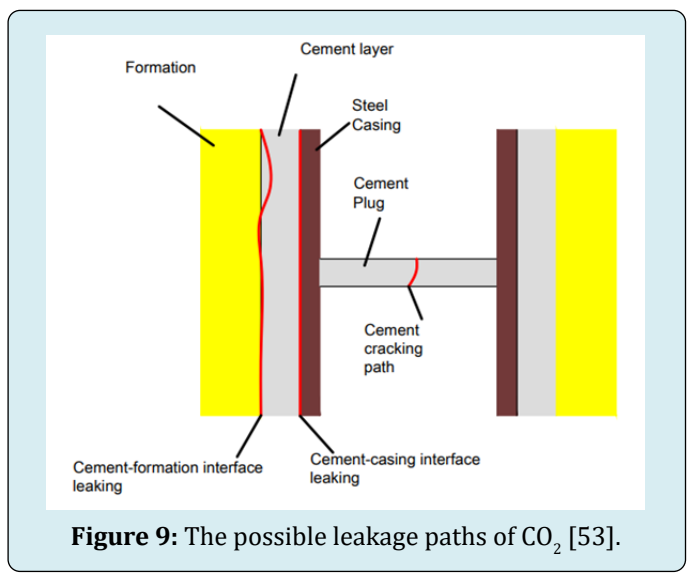

**Figure 9:** The possible leakage paths of $CO_2$ [53].

**Estimated Time for Carbonation Process:** Many research groups have predicted the carbonation times of cement exposed to $CO_2$ by using experimental and mathematical models. The carbonation is a substantial variable to evaluate the quality of the Carbon Capture Storage. Therefore, it is essential to obtain it. The experimental and numerical methods have their advantages and limitations. The empirical model in some situations will produce inaccurate results due to equipment error, and occasionally it is hard to manage the procedure properly due to external conditions beyond our observation, and most likely, the cost to do experiments is more expensive and time-consuming. In contrast, on average, the numerical model is much easier to do. It will eliminate the errors from equipment and human. It will produce the result faster. The most important thing for using a numerical model is to set up a governing equation with correct initial and boundary conditions. Although showing the limitation on both experimental and numerical models, they are used to verify one another. That is a reason why it is necessary to study both.

**Experimental Measurements:** Duguid A [53] designed an experiment to forecast the time to deteriorate the cement sheath in a well exposed to carbonated brine. The samples were created by drilling the stone cylinder 55 mm in diameter, 10 mm in height with a 25-mm axial hole. The minimum depth from the outside of the cylinder to the boundary of the hole was 3mm. According to Duguid A [53], the depth of reaction was quantified at five different locations: 0, 45, 90,135,180 degrees as presented in Figure 10 in order to evaluate how different the surface cement would react with carbonate brine.

The relationship between carbonation depth and time$^{1/2}$/ radius is linear in most cases. The most linear relationship is at the condition pH=3 and T=50 °C, and this condition also causes the most carbonation for cement. It is in agreement with Duguid A, et al. [54]. The prediction time for 25 mm cement sheath to be deteriorated is approximately from 30,000 to 70,000 years if the favorable cement is selected and the good cementing job is done.

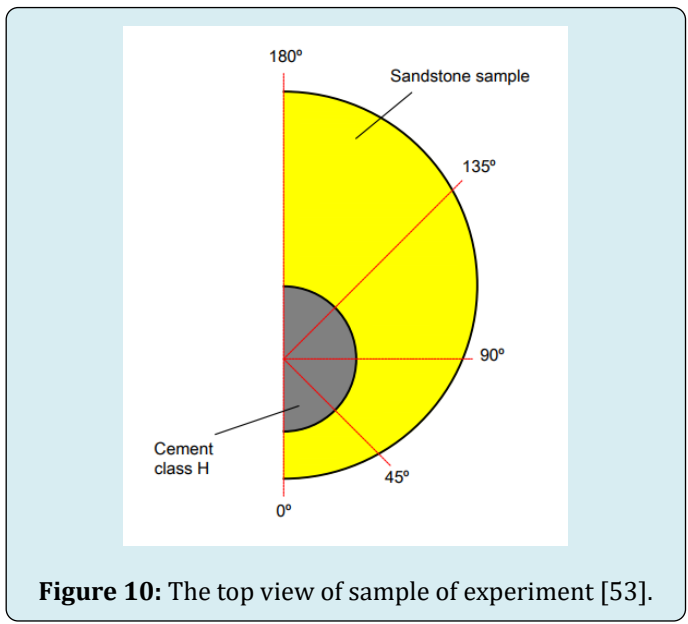

**Figure 10:** The top view of sample of experiment [53].

Kutchko BG [55] showed the reaction rate when the cement was exposed to supercritical $CO_2$ and $CO_2$-saturated brine. Supercritical $CO_2$ is a separate free phase causing hydrodynamic trapping. On the other hand, some would dissolve in the brine existing in $CO_2$ saturated brine form, which induces the solubility trapping. The cement samples were embedded in 1% NaCl at 30.3 MPa and 50 °C under static conditions. The estimation for penetration depth is 1±0.07 mm for the $CO_2$-saturated brine and 2.9±0.89 mm for the supercritical $CO_2$ after 30 years. It indicated that the supercritical $CO_2$ would degrade Portland cement faster than the $CO_2$ saturated brine. This is comprehensible because the condition of supercritical $CO_2$ is at high pressure and high temperature. The penetration depth over time for the $CO_2$-saturated brine and supercritical can be found in Figure 11.

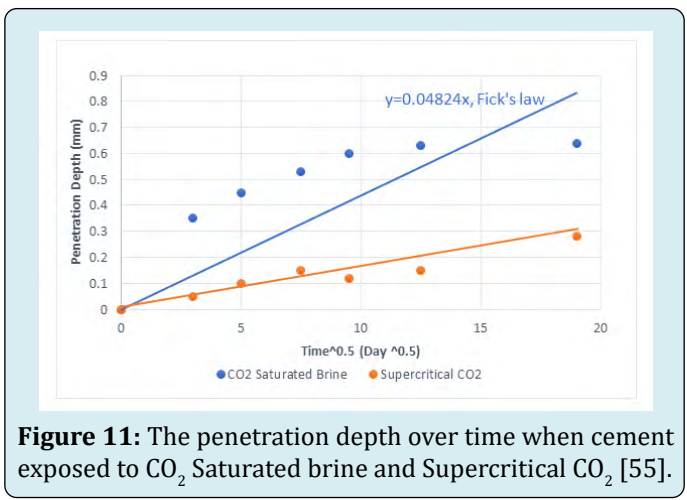

**Figure 11:** The penetration depth over time when cement exposed to $CO_2$ Saturated brine and Supercritical $CO_2$ [55].

Zhang L, et al. [56] introduced another approach to estimate the penetration depth over time. Fick's diffusion and Elovich's equation were fit to experimental data. Elovich's equation (Equation 16) can be shown by Allen JA, et al. [57] and Kutchko BG, et al. [58].

$$\frac{dL}{dt} = a * exp\left(-bL\right) \quad (16)$$

Where L is the penetration depth(mm) at time t (days) of exposure and a, b are constants decided from experiment data.

Integrating the Equation 16 above with respect to t yields Equation 17:

$$L = \frac{1}{b}ln(t) + \frac{1}{b}ln\left(ab\right) \quad (17)$$

Where a, b can be estimated to fit the data a=2.47, b=22.08

Estimation of penetration depth with Elovich's equation is more accurate than Fick's diffusion as shown in Figure 12. Elovich's equation has been used in several kinetic studies [59-61]. However, the outcomes reaffirmed Elovich's equation as a powerful method to measure the $CO_2$ penetration depth.

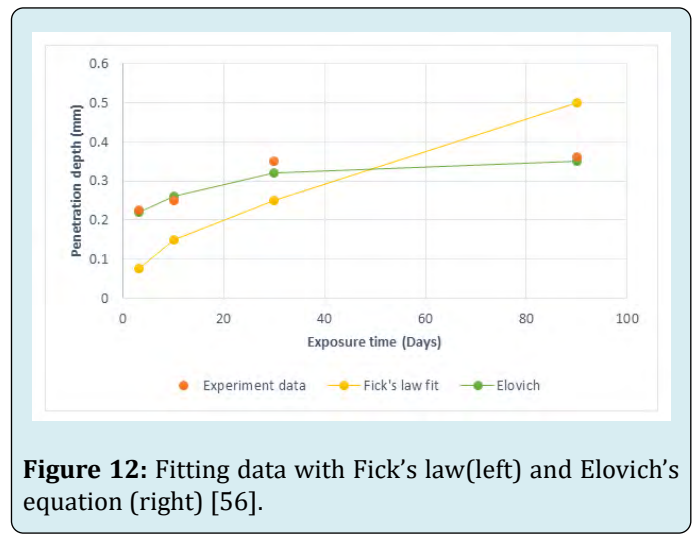

**Figure 12:** Fitting data with Fick's law(left) and Elovich's equation (right) [56].

In Duguid A, et al. [54], the experiment for limestone and sandstone- like condition were executed. Class H cement pastes were exposed to temperatures from 20 to 50 °C and pH 2.4 to 5. Then the samples were interpreted by using multiple techniques such as Inductively Coupled Plasma Optical Emission Spectroscopy (ICP-OES), optical microscopy, X-ray diffraction, and Electron Probe Microanalysis (EPMA). The experimental model was designed as presented in Figure 13. The $CO_2$ air was percolated into the carbonated brine and then pumped to the reactor vessel, which contains the cement sample. The purpose of the recirculated flow is to assure that it was saturated with $CaCO_3$ before reaching the reaction vessel. No observable degradation in the limestone-like condition was observed. Under the sandstone-like condition, there are 5 distinct layers that appeared: orange, brown, white, gray, and core. Each layer would expose the different behavior to carbonated brine. The orange and brown part display a leached region. The white layer shows a carbonated region. The gray section depicts a calcium hydroxide dissolution region, and the core section is no change. The outer layer was degraded fully at pH=2.4, 3.7, and temperatures 20℃ and 50℃. The sample got the most damage at pH=2.4 and T=20℃, and the least degradation occurred at pH= 5 and T=50℃. This conclusion is illustrated by the dissolution of carbon dioxide in water. The carbon dioxide solubility in water increases with decreasing temperature, and more carbonic acid is formed. Therefore, the most degradation was occurred at pH=2.4 and temperature 20 ℃.

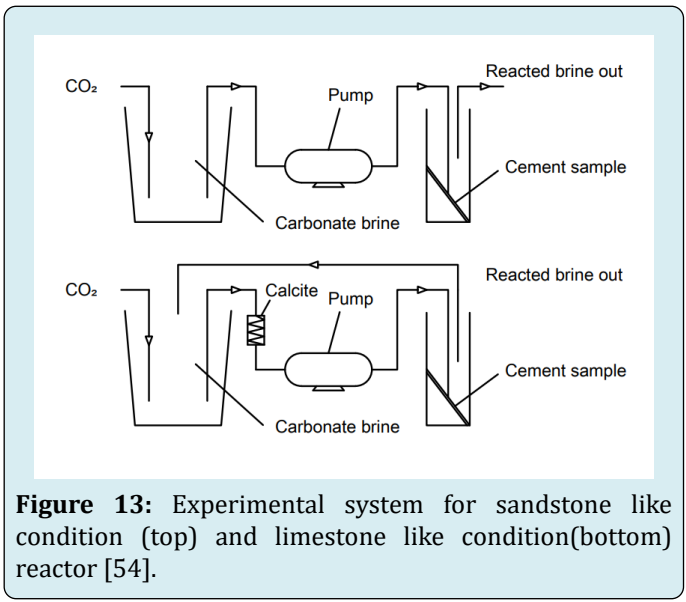

**Figure 13:** Experimental system for sandstone like condition (top) and limestone like condition(bottom) reactor [54].

Carey JW, et al. [62] and Carey JW, et al. [63] investigated the behavior of wellbore integrity and $CO_2$-brine flow along the casing-cement micro annulus. The core-flood examination was performed at 40℃, 14 MPa pore pressure, and 28 MPa confining pressure. The experimental system included a 10 cm length of limestone with a combination of rectangular steel embedded in the cement. The blended solution, 50% supercritical $CO_2$ and 50% brine, was run through limestone/cement combination. There are two corrosion processes: steel corrosion and cement corrosion. The corrosion on steel occurs by the electrochemical mechanism. The film $FeCO_3$ layer was formed from the $CO_2$-rich fluid in contact with steel. The film layer protected the steel from deeper penetration of the $CO_2$-rich fluid. However, the solubility of the $FeCO_3$ layer would increase if the pH decreases. As a result, corrosion rate is increased, with increasing flow rate of the $CO_2$-rich fluid. For cement degradation, the rate is dependent on cement properties and the flow rate of $CO_2$-rich fluid. The diffusion coefficient was discovered in the interval from 10-12 to 10-10 cm2/sec by assuming a 1D diffusion problem with characteristic diffusion time 2√Dt and penetration depth from 50-250µm.

Adeoye JT, et al. [64] carried out experiments of a novel engineered cementitious composite (ECC) exposed to $CO_2$-saturated water under static and flow conditions at 10 MPa and 50℃. The depth of alteration estimated was 72 mm over 50 years by Fick's law. In a similar study was performed by Kutchko BG, et al. [65] at 15 MPa and 50℃, the depth of alteration was 224 mm over 50 years. The higher pressure of CO2 in the study Kutchko BG, et al. [65] was not probably the main reason to lead to a 3 three times increase of carbonation depth. Nonetheless, the exciting finding lay on the pozzolan to cement ratio. The material used in the study Kutchko BG, et al. [65] has the pozzolan to cement ratio of 65:35, while this ratio in the study Adeoye JT, et al. [64] is 45:55. Thus, the increase of the pozzolan would lead to faster carbonation.

**Mathematical Prediction:** Along with experimental works, there are very few studies using the robust mathematical model to predict the penetration depth during geological storage $CO_2$. The mathematical model may accompany experimental work, which will conserve resources compared to purely experimental work. Below are some remarkable studies which have been done so far.

Tao Q, et al. [66,67] developed a mathematical model to investigate some relationships during geological $CO_2$ storage. The leaking pathway of $CO_2$ includes two parts, as shown in Figure 14a: the bottom due to cement degradation, the top is the water-saturated porous medium where an assumption of no resistance was made. Figure 14b showed how the pressure of the reservoir changes during $CO_2$ injection. The pressure of the reservoir increased as $CO_2$ was injected and decreased after the injection was stopped.

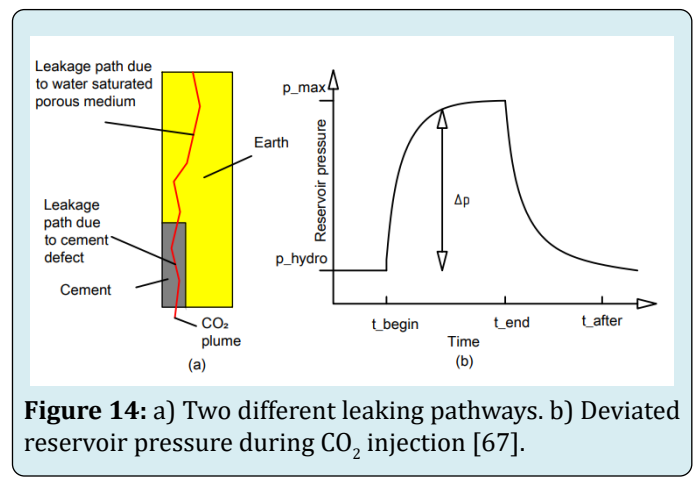

**Figure 14:** a) Two different leaking pathways. b) Deviated reservoir pressure during $CO_2$ injection [67].

If there is only buoyancy force causing $CO_2$ flow, the potential gradient of $CO_2$ is calculated by Equation 18:

$$\nabla \varphi = \nabla(\Delta \rho g z) \quad (18)$$

Where $\Delta \rho$ is the density difference between $H_2O$ and $CO_2$ z and g are the depth and gravitational, respectively.

During the injection period, both buoyancy and pressure elevation contribute to drive $CO_2$. Therefore, the potential gradient of $CO_2$ is computed by Equation 19:

$$\nabla \varphi = \nabla(\Delta \rho g z) + \nabla p_c \quad (19)$$

Where $p_c$ is the capilary pressure.

Several wells were investigated during the geological $CO_2$ storage. The $CO_2$ flux was discovered to be responsive to pressure elevation and the leakage depth. The $CO_2$ flux decreases while increasing the leakage depth due to the increasing of $CO_2$ density. The relationship between the $CO_2$ leakage flux and injection pressure at shallow leak (4000 ft) and deep leak (10000 ft) was investigated. The $CO_2$ leakage flux increases with increasing injection pressure, but the rates are different. The $CO_2$ leakage flux rate is more at deep leak than it does at the shallow leak. According to the author, this is because the injection pressure would overcome the buoyancy pressure at a shallow leak. In contrast, at a deep leak, the buoyancy pressure decreases more gradually due to the increasing $CO_2$ density.

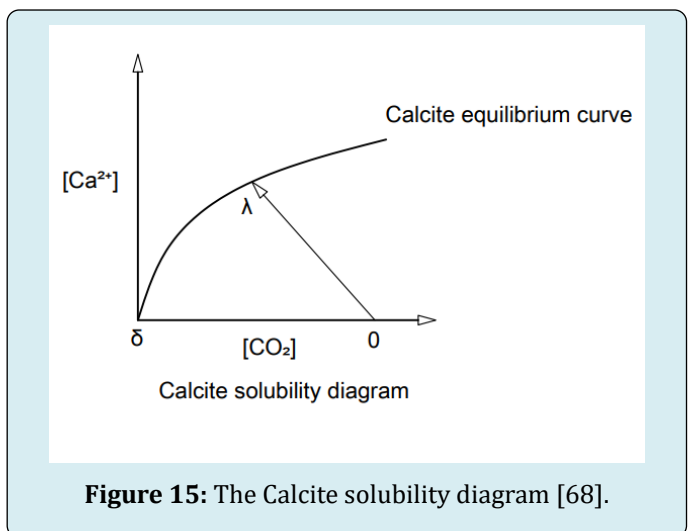

**Figure 15:** The Calcite solubility diagram [68].

Deremble L, et al. [68] simulated the evolution of layers under $CO_2$-rich brine flow. The development of layer calcite or silica gel will be proportional, while the flux of calcium or $CO_2$ is inversely proportional to the square root of time. An increase in $CO_2$ flow rate increases calcium dissolution rate until calcite equilibrium is reached at point $\lambda$, as presented in Figure 15. Then ion calcium cannot be discharged anymore. As a result, $CO_2$ would react with other species in the cement until it approaches point δ where all species have gone, and then the mathematical model takes into consideration the additional physical aspects including: micro-annulus geometry, the Peclet number, and the characteristic length scales of a defect. The model uses the implicit algorithm to solve for a solution. It also affirms that the penetration depth is proportional to the square root of time.

Huet BM, et al. [69] connected the geochemical and transport module to simulate the degradation of cement during geological $CO_2$ storage. The Dynaflow was adopted to solve a non-linear system of partial differential equations. Both Galerkin finite element and vertex centered finite volume space discretization of transport equations were executed. An implicit backward finite difference time stepping of the transport equation was applied to produce the results. The effective diffusion coefficient needs to be assumed in order of 10-11 $m^2/s$ by Bentz DP, et al. [70]. The thickness of the calcite layer is also proportional to the square root of time. The difference between experimental data and the model occurred due to the diffusion coefficient estimated. The growth of calcite carbonate concentration over time as presented in Figure 16 would provide insight into the transport of $CO_2$ to cement. The carbonate concentration increases very rapidly while its radius decreases with time elapsed. There are two distinct regions with different transport mechanisms. In the first region with time exposure less than 60 days, carbonate species ($CO_2$, $HCO_3^-$) disperse into the sample through the calcite and the silica gel layer. The second region is where the calcite layer dissolves.

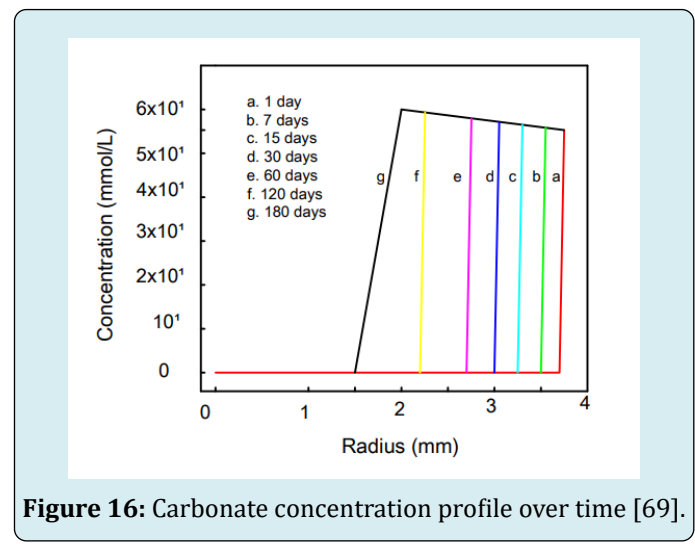

**Figure 16:** Carbonate concentration profile over time [69].

Also, well integrity has been investigated extensively in Hawkes C, et al. [71]; Hawkes CD, et al. [72]; Scherer GW, et al. [73]; Neuville N, et al. [74] to evaluate for long-term $CO_2$ storage. The downhole testing programs, such as cement sheath pressure transient testing, mini-frac testing, cement sampling, and fluid sampling, were performed.

## Outlook

The traditional Portland cement carries many advantages such as low cost, high compressive strength, low alkali content, and long-term stability. Nevertheless, cement is sensitive in an acidic environment, and the cement industry is a source of $CO_2$ emissions. Therefore, Portland cement is not the most excellent sealant material to serve in CCS project. Based on the nature of CCS project, the most optimal cement could prevent corrosion under the acidic environment. Also, the cement should possess low permeability, porosity, and high mechanical strength.

Mahmoud AA, et al. [75] proposed to add Synthetic Polypropylene Fiber (PPF) into Class G cement to improve it. Four samples with 0%(PPF0), 0.125%(PPF1), 0.25%(PPF2), 0.375%(PPF3) of PPF were arranged for the experiment. The results of this study showed that the carbonation depth and carbonation rate decreased while compressive strength and tensile strength increased, as presented in Figures 17-20, respectively. The decreasing of carbonation depth and carbonation rate indicated the reduction of cement permeability.

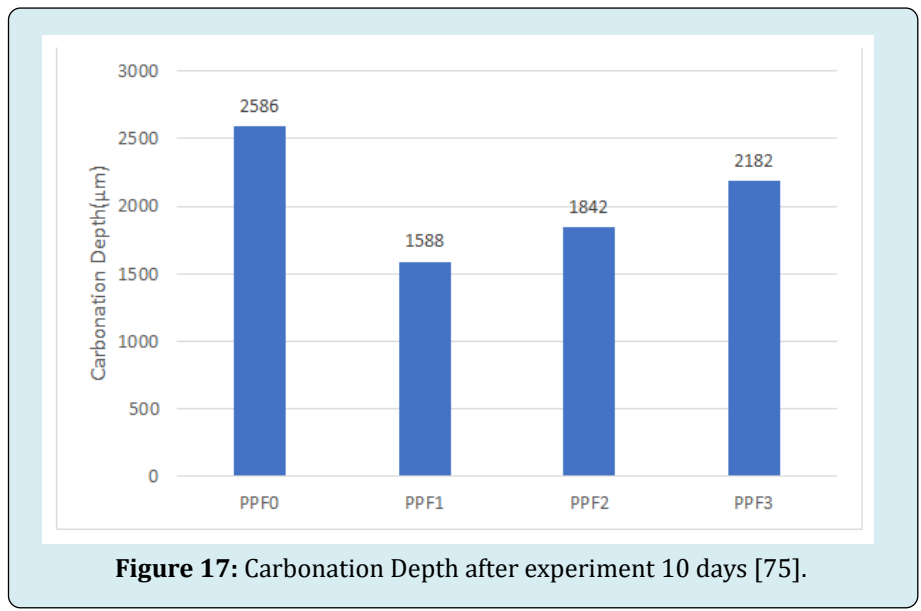

**Figure 17:** Carbonation Depth after experiment 10 days [75].

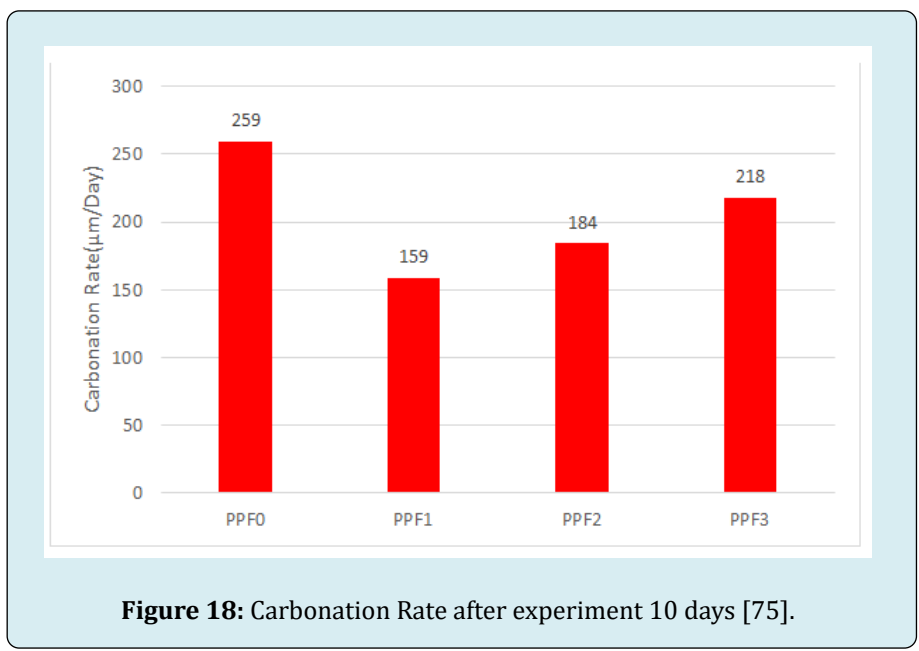

**Figure 18:** Carbonation Rate after experiment 10 days [75].

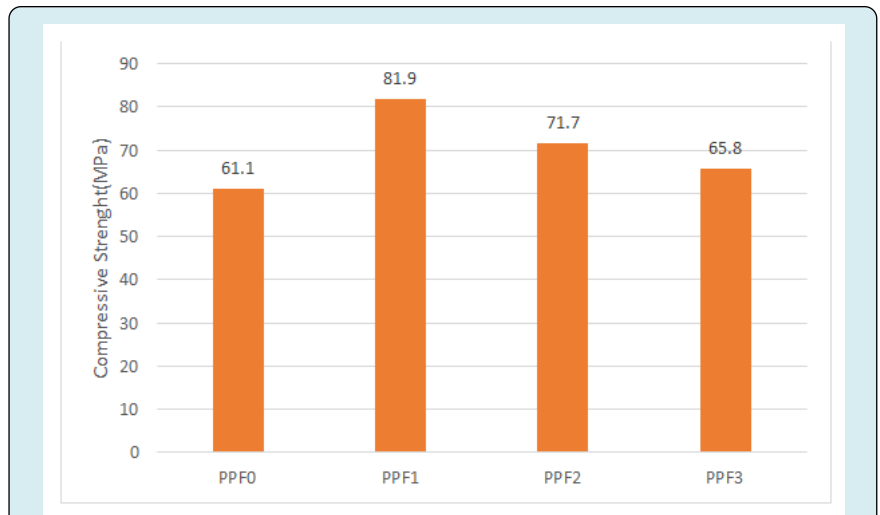

**Figure 19:** Compressive strength after experiment 10 days [75].

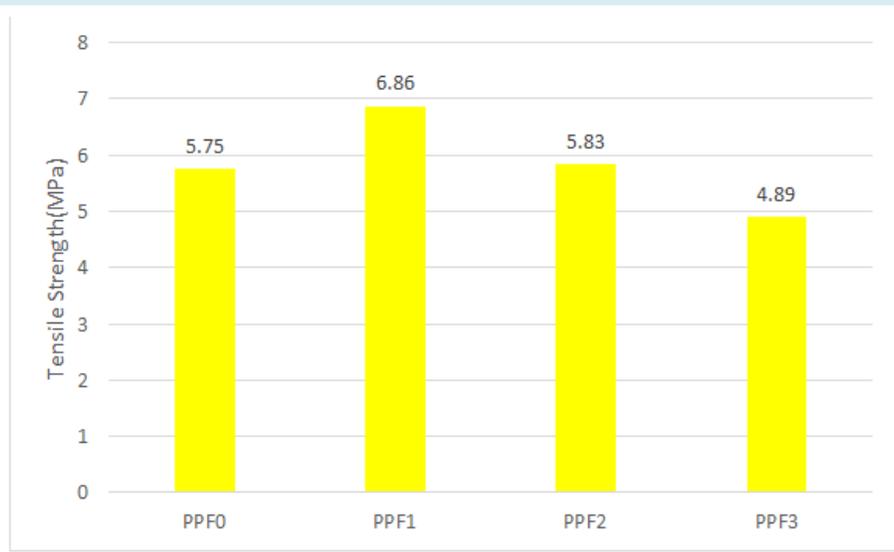

**Figure 20:** Tensile Strength after experiment 10 days [75].

Nanomaterial such as nano-silica (nano $SiO_2$) [8], nano-alumina (nano $Al_2O_3$) [76], nano-titanium dioxide (nano$TiO_2$) [9], carbon nanotubes (CNTs) [77], Polymer/clay nanocomposites [78], nanoglass flake (NGFS) [79] were considered as good additive to improve cement quality because of their large surface area and reactivity. The nanomaterials can solidify the cement microstructure and lessen the porosity, then advance the mechanical strength.

Ponzi GGD, et al. [80] proposed basalt powder as an additive material in cement formulation because the basalt powder has low pozzolanic activity, large inert fraction, and small particle size. The basalt power plays a role as a filling-substance to the porous cement networks to curb the fluid intrusion. The experimental results explored that the formulation with low basalt powder content (≤ 0.5 w.%) exhibited more resistance to $CO_2$ degradation, lower porosity and permeability, and stronger mechanical properties.

Other sealant materials can replace traditional Portland cement, such as geopolymer cement, resins, biofilms barriers, foams [81]. Geopolymer, such as zeolites, was discovered to have better resistance with $CO_2$- rich brine because it contains less calcium oxide than Portland cement does. Resins are particle-free fluids with low mobility, hard, rigid, and impermeable materials. They include phenolic, epoxy, and furan resins. Biofilm sealants include urea, $Ca^{2+}$, nutrient feed, and micro-organism. The principle is to accelerate calcium to form calcite and seal fractures. Foam is a gas-liquid blend, and it can block the flow rate of $CO_2$ in porous media and increase the $CO_2$ viscosity.

Because carbon capture and storage projects are expanding, so the cement consumption is increasing. Those materials introduced above have many advantages, but the disadvantages still exist. Therefore, the new research direction should focus on improving the cement quality. For example, geopolymer is detrimental to human health. Thus, new components should be studied to make it become human friendly.

## Conclusions

This study's major goal was to review the previous papers for carbon capture and storage project. Many experiments have been performed to assess the well integrity and predict the time degradation of geological $CO_2$ storage, but there are very few mathematical models. Research predicts 30,000-70,000 years for 25 mm cement to be carbonated. Some studies tried to match the experimental data to a particular equation and introduce that the penetration depth is proportional to the square root of time. Furthermore, the debonding interface between casing/cement or cement/formation is the primary cause for leakage of $CO_2$. However, it has not been thoroughly investigated. Some substantial conclusions can be drawn from this review:

- The need for a more accurate mathematical model to evaluate the well integrity and anticipate the corrosion rate during geological $CO_2$ storage is very crucial.
- A further investigation should identify the debonding interface between casing/cement or cement/formation issue and predict how the micro-annuli of case/cement or cement/formation behaves with the variation of temperature, stress, and chemical reactions during geological $CO_2$ storage.
- The diffusion coefficient is one of the most crucial parameters in the corrosion process. However, it has not been studied sufficiently in petroleum corrosion. Hence, it should be an interesting topic for future studies.
- Improving the cementing property is one of the means to curb the corrosion rate. Future studies should investigate more how to reduce reactive species and add more inhibitors to advance Portland cement quality.

If these tasks are done properly, it will clear up a considerable concern to make the operation more predictable and administered.

## Author Contribution

Writing- original draft preparation, Nguyen V; writing-review and editing, Nguyen V, Olatunji O, Guo B and Ning Liu. All authors have read and agreed to the published version of the manuscript.

## Funding

This research received no external funding

## Conflicts of Interest

The authors declare no conflict of interest.

## References

1. Zhang M, Bachu S (2011) Review of integrity of existing wells in relation to $CO_2$ geological storage : What do we know ?. Int J Greenh Gas Control 5(4): 826-840.

2. Young JF (2001) Portland Cements. In: Buschow KJ, Cahn RW, Flemings MC, Ilschner B, Kramer EJ, et al. (Eds.), Encyclopedia of Materials: Science and Technology. Elsevier, Oxford, USA, pp: 7768-7773.

3. MacLaren DC , White MA (2003) Cement: Its Chemistry and Properties. J Chem Educ 80(6): 623-635.

4. Duguid, A (2006) The Effect of Carbonic Acid on Well Cements. Doctoral Thesis, Princeton University, NJ.

5. Duguid A, Radonjic M, Scherer GW (2011) Degradation of cement at the reservoir/cement interface from exposure to carbonated brine. Int J Greenh Gas Control 5(6): 1413-1428.

6. Lee BY, Kurtis KE, Jayapalan AR (2013) Effects of nano-$TiO_2$ on properties of cement-based materials. Mag Concr Res 65(21): 1293-1302.

7. Sobolev K (2015) Nanotechnology and Nanoengineering of Construction Materials in Construction. In: Sobolev K, Shah SP (Eds.), Nanotechnology Construction. Springer: Berlin, Germany, pp: 3-13.

8. Choolaei M, Rashidi AM, Ardjmand M, Yadegari A, Soltanian H (2012) The effect of nanosilica on the physical properties of oil well cement. Mater Sci Eng A 538: 288-294.

9. Chen J, Kou SC, Poon CS (2012) Hydration and properties of nano-$TiO_2$ blended cement composites. Cem Concr Compos 34(5): 642-649.

10. Nasvi MCM, Ranjith PG, Sanjayan J (2013) The permeability of geopolymer at down-hole stress conditions: Application for carbon dioxide sequestration wells. Appl Energy 102: 1391-1398.

11. Andre MA (2008) Mathematical Modeling and Multiscale Simulation of $CO_2$ Storage in Saline Aquifers. Doctoral Thesis, Stanford University, USA, pp: 1-207.

12. Martinez ASA (2017) Mathematical Model for Predicting the Carbon Sequestration Potential of Exposed Ordinary Portland cement (OPC) concrete. Master Thesis, University of Colorado, USA.

13. Duguid A, Guo B, Nygaard R (2017) Well Integrity Assessment of Monitoring Wells at an Active CO2-EOR Flood. Energy Procedia 114: 5118–5138.

14. Skocek J, Zajac M, Haha MB (2020) Carbon Capture and Utilization by mineralization of cement pastes derived from recycled concrete. Sci Rep 10: 1-12.

15. Wise JLEE (2019) Wellbore Integrity and Cement Sheath. Masters Thesis, Oklahoma State University, USA, pp: 1-85.

16. Orlic B, Heege J, Wassing B (2011) Assessing the integrity of fault- and top seals at $CO_2$ storage sites. Energy Procedia 4: 4798-4805.

17. Hofstee C, Seeberger F, Orlic B, Mulders F, Van Bergen F, et al. (2008) The feasibility of effective and safe carbon dioxide storage in the De Lier gas field. First Break 26: 53-57.

18. Orlic B (2009) Some geomechanical aspects of geological $CO_2$ sequestration. KSCE J Civ Eng 13: 225-232.

19. Li B, Li H, Zhou F, Guo B, Chang X (2017) Effect of cement sheath induced stress on well integrity assessment in carbon sequestration fields. J Nat Gas Sci Eng 46: 132-142.

20. Omosebi O, Ahmed R, Shah S, Osisanya S (2015) Mechanical integrity of well cement under geologic carbon sequestration conditions. Proceeding of annual Carbon Management Technology Conference, Texas, USA,

21. Shadravan A, Schubert J, Amani M, Teodoriu C (2015) Using fatigue-failure envelope for cement-sheath-integrity evaluation. SPE Drill Complet 30: 68-75.

22. Teodoriu C, Kosinowski C, Amani M, Schubert J, Shadravan A (2013) Wellbore integrity and cement failure at hpht conditions. Int J Eng Appl Sci 2: 1-13.

23. Boukhelifa L, Moroni N, James SG, Le Roy Delage S, Thiercelin MJ, et al. (2004) Evaluation of cement systems for oil and gas well zonal isolation in a full-scale annular geometry. Proc Drill Conf pp: 825-839.

24. Todorovic J, Gawel K, Lavrov A, Torsæter M (2016) Integrity of downscaled well models subject to cooling. SPE Bergen One Day Seminar, Grieghallen, Bergen, Norway.

25. Aursand P, Hammer M, Lavrov A, Lund H, Munkejord ST, et al. (2017) Well integrity for CO2 injection from ships: Simulation of the effect of flow and material parameters on thermal stresses. Int J Greenh Gas Control 62: 130-141.

26. Lund H, Torseter M, Munkejord ST (2016) Study of thermal variations in wells during carbon dioxide injection. SPE Drill Complet 31: 159-165.

27. Ruan B, Xu R, Wei L, Ouyang X, Luo F, et al. (2013) Flow and thermal modeling of $CO_2$ in injection well during geological sequestration. Int J Greenh Gas Control 19: 271-280.

28. Lavrov A, Todorovic J, Torsaeter M (2015) Numerical study of tensile thermal stresses in a casing-cement-rock system with heterogeneities. Am Rock Mech Assoc 4: 2996-3004.

29. Engineering Tool Box (2008) Concrete-Properties.

30. Thiercelin MJ, Dargaud B, Baret JF, Rodriquez WJ (1998) Cement design based on cement mechanical response. SPE Drill Complet 4: 266-273.

31. Hangx SJT, Spiers CJ, PeachCJ (2010) Mechanical behavior of anhydrite caprock and implications for CO2 sealing capacity. J Geophys. Res. Solid Earth 115(B7): 1-22.

32. Kim K, Vilarrasa V, Makhnenko RY (2018) $CO_2$ injection effect on geomechanical and flow properties of calcite-rich reservoirs. Fluids 3(3): 66.

33. Charalampidou EM, Garcia S, Buckman J, Cordoba P, Lewis H, et al. (2017) Impact of CO2-induced Geochemical Reactions on the Mechanical Integrity of Carbonate Rocks. Energy Procedia 114: 3150-3156.

34. Luquot L, Gouze P (2009) Experimental determination of porosity and permeability changes induced by injection of $CO_2$ into carbonate rocks. Chem Geol 265(1-2): 148-159.

35. Rohmer J, Pluymakers A, Renard F (2016) Mechano-chemical interactions in sedimentary rocks in the context of $CO_2$ storage: Weak acid, weak effects?. Earth-

Science Rev 157: 86-110.

36. Bemer E, Lombard JM (2010) From Injectivity to Integrity Studies of $CO_2$ Geological Storage: Chemical Alteration Effects on Carbonates Petrophysical and Geomechanical Properties. Oil Gas Sci Technol 65: 445-459.

37. Vanorio TV, Nur A, Ebert Y (2011) Rock physics analysis and time-lapse rock imaging of geochemical effects due to $CO_2$ injection into reservoir rocks. Geophysics 76(5): 23-33.

38. Iyer J, Chen X, Carroll SA (2020) Impact of Chemical and Mechanical Processes on Leakage from Damaged Wells in $CO_2$ Storage Sites. Environ Sci Technol 54: 1196-1203.

39. Dávila G, Cama J, Chaparro MC, Lothenbach B, Schmitt DR, et al. (2021) Interaction between CO2-rich acidic water, hydrated Portland cement and sedimentary rocks: Column experiments and reactive transport modeling. Chem Geol 572.

40. Tang Y, Hu S, He Y, Wang Y, Wan X, et al. (2021) Experiment on $CO_2$-brine-rock interaction during $CO_2$ injection and storage in gas reservoirs with aquifer. Chem Eng J 413.

41. Santra A, Sweatman R (2011) Understanding the Long-Term Chemical and Mechanical Integrity of Cement in a CCS Environment. Energy Procedia 4: 5243-5250.

42. Ojala IO (2011) The effect of $CO_2$ on the mechanical properties of reservoir and cap. Energy Procedia 4: 5392-5397.

43. Morris JP, Hao Y, Foxall W, Mcnab W (2011) A study of injection-induced mechanical deformation at the In Salah CO 2 storage project. Int J Greenh Gas Control 5: 270-280.

44. Karimnezhad M, Jalalifar H, Kamari M (2014) Investigation of caprock integrity for $CO_2$ sequestration in an oil reservoir using a numerical method. J Nat Gas Sci Eng 21: 1127-1137.

45. Wolterbeek TKT, Peach CJ, Spiers CJ (2013) Reaction and transport in wellbore interfaces under $CO_2$ storage conditions : Experiments simulating debonded cement–casing interfaces. Int J Greenh Gas Control 19: 519-529.

46. Gray LGS, Choi Y, Young D, Ne S (2013) Wellbore integrity and corrosion of carbon steel in CO 2 geologic storage environments : A literature review. Int J Greenh Gas Control 16(S1): 70-77.

47. Voormeij DA, Simandl GJ (2002) Geological and Mineral CO2 Sequestration Options: A technical review. Br Columbia Geol Surv Geol Fieldwork, pp: 265-277.

48. Russick EM, Poulter GA, Adkins CLJ, Sorensen NR (1996) Corrosive effects of supercritical carbon dioxide and cosolvents on metals. J Supercrit Fluids 9(1): 43-50.

49. Seiersten M, Kongshaug KO (2005) Materials Selection for Capture, Compression, Transport and Injection of $CO_2$. Carbon Dioxide Capture Storage Deep Geol Form 2: 937-953.

50. Seiersten M (2001) Material Selection for Separation, Transportation and Disposal of $CO_2$. Corrosion 2001, NACE, Houston, TX, USA.

51. Choi YS, Nesic S, Young D (2010) Effect of impurities on the corrosion behavior of CO2 transmission pipeline steel in supercritical $CO_2$-water environments. Environ Sci Technol 44: 9233-9238.

52. Lin G, Zheng M, Bai Z, Zhao X (2006) Effect of temperature and pressure on the morphology of carbon dioxide corrosion scales. Corrosion 62: 501-507.

53. Duguid A (2009) An estimate of the time to degrade the cement sheath in a well exposed to carbonated brine. Energy Procedia 1: 3181-3188.

54. Duguid A, Scherer GW (2010) Degradation of oil well cement due to exposure to carbonated brine. Int J Greenh Gas Control 4: 546-560.

55. Kutchko BG (2008) Effect of C0$_2$ on the Integrity of Well Cement under Geologic Sequestration Conditions. Doctoral Thesis, Carnegie Mellon University, Pennsylvania, USA.

56. Zhang L, Dzombak DA, Nakles DA, Hawthorne SB, Miller DJ, et al. (2014) Rate of H 2 S and CO 2 attack on pozzolan-amended Class H well cement under geologic sequestration conditions. Int J Greenh Gas Control 27: 299-308.

57. Allen JA, Scaife PH (1966) The Elovich equation and chemisorption kinetics. Aust J Chem 19.

58. Kutchko BG, Strazisar BR, Lowry GV, Dzombak DA, Thaulow, et al. (2008) Rate of $CO_2$ attack on hydrated class H well cement under geologic sequestration conditions. Environ Sci Technol 42: 6237-6242.

59. Juang RS, Chen ML (1997) Application of the Elovich Equation to the Kinetics of Metal Sorption with Solvent-Impregnated Resins. Ind Eng Chem Res 36: 813-820.

60. Wu FC, Tseng RL, Juang RS (2009) Characteristics of Elovich equation used for the analysis of adsorption kinetics in dye-chitosan systems. Chem Eng J 150(2-3): 366-373.

61. Aharoni C, Suzin Y (1982) Application of the Elovich equation to the kinetics of occlusion. Part 1.-Homogeneous microporosity. J Chem Soc 78: 2321-2327.

62. Carey JW, Svec R, Grigg R, Zhang J, Crow W (2010) Experimental investigation of wellbore integrity and CO2 – brine flow along the casing – cement microannulus. Int J Greenh Gas Control 4: 272-282.

63. Carey JW, Svec R, Grigg R, Lichtner PC, Zhang J (2009) Well bore integrity and $CO_2$ -brine flow along the casing-cement microannulus. Energy Procedia 1: 3609-3615.

64. Adeoye JT, Beversluis C, Murphy A, Li VC, Ellis BR (2019) Physical and chemical alterations in engineered cementitious composite under geologic CO2 storage conditions. Int J Greenh Gas Control 83: 282-292.

65. Kutchko BG, Strazisar BR, Huerta N, Lowry GV, Dzombak DA, et al. (2009) $CO_2$ reaction with hydrated class H well cement under geologic sequestration conditions: Effects of flyash admixtures. Environ Sci Technol 43(10): 3947-3952.

66. Tao Q, Checkai D, Huerta N, Bryant SL (2010) Model to Predict $CO_2$ Leakage Rates Along a Wellbore. In Proceeding of the SPE Annual Technical Conference and Exhibition, Italy.

67. Tao Q, Checkai D, Huerta N, Bryant SL (2011) An Improved Model to Forecast $CO_2$ Leakage Rates Along a Wellbore. Energy Procedia 4: 5385-5391.

68. Deremble L, Loizzo M, Huet B, Lecampion B, Quesada D (2011) Stability of a leakage pathway in a cemented annulus. Energy Procedia 4: 5283-5290.

69. Huet BM, Prevost JH, Scherer GW (2010) Quantitative reactive transport modeling of Portland cement in CO2-saturated water. Int J Greenh Gas Control 4: 561-574.

70. Bentz DP, Jensen OM, Coats, AM, Glasser FP (2000) Influence of silica fume on diffusivity in cement-based materials. I. Experimental and computer modeling studies on cement pastes. Cem Concr Res 30: 953-962.

71. Hawkes C, Gardner C, Watson T, Chalaturnyk R (2011) Overview of wellbore integrity research for the IEA GHG Weyburn-Midale CO 2 Monitoring and Storage Project. Energy Procedia 4: 5430-5437.

72. Hawkes CD, Gardner C (2013) Pressure transient testing for assessment of wellbore integrity in the IEAGHG Weyburn – Midale $CO_2$ Monitoring and Storage Project. Int J Greenh Gas Control 16(S50-S61).

73. Scherer GW, Kutchko BG, Thaulow N, Duguid A, Mook B (2011) Characterization of cement from a well at Teapot Dome Oil Field : Implications for geological sequestration. Int J Greenh Gas Control 5: 115-124.

74. Neuville N, Aouad G, Lecolier E, Damidot D (2012) Innovative Leaching Tests of an Oilwell Cement Paste for $CO_2$ Storage : Effect of the Pressure at 80°C. Energy Procedia pp: 472-479.

75. Mahmoud AA, Elkatatny S (2020) Improving class G cement carbonation resistance for applications of geologic carbon sequestration using synthetic polypropylene fiber. J Nat Gas Sci Eng pp: 76.

76. Oltulu M, Şahin R (2011) Single and combined effects of nano-SiO2, nano-Al2O3 and nano-Fe2O3 powders on compressive strength and capillary permeability of cement mortar containing silica fume. Mater Sci Eng A 528: 7012-7019.

77. Nasibulina LI, Anoshkin IV, Shandakow SD, Nasibulin AG, Cwirzen A, et al. (2010) Direct synthesis of carbon nanofibers on cement particles. Transp Res Rec 2142(1): 96-101.

78. Hakamy A, Shaikh FUA, Low IM (2014) Thermal and mechanical properties of hemp fabric-reinforced nanoclay-cement nanocomposites. J Mater Sci 49: 1684-1694.

79. Salehi S, Ehsani M, Khonakdar HA (2017) Assessment of thermal, morphological, and mechanical properties of poly(methyl methacrylate)/glass flake composites. J Vinyl Addit Technol 23(1): 62-69.

80. Ponzi GGD, Santos VHJ, Martel RB, Pontin D, Stepanha ASDGE, et al. (2021) Basalt powder as a supplementary cementitious material in cement paste for CCS wells: chemical and mechanical resistance of cement formulations for CO2 geological storage sites. Int J Greenh Gas Control pp: 109.

81. Zhu D, Peng S, Zhao S, Wei M, Bai B (2021) Comprehensive Review of Sealant Materials for Leakage Remediation Technology in Geological CO2Capture and Storage Process. Energy and Fuels 35: 4711-4742.